# Blockchain-based RBAC Model with Separation of Duties constraint in Cloud Environment

Ok-Chol Ri [1], Yong-Jin Kim [2], and You-Jin Jong [1]

[1]Kum Sung Middle School Number 2, PyongYang, 999093, D.P.R of Korea
[2]Faculty of Mathematics, KIM IL SUNG University, PyongYang, 999093, D.P.R of Korea

Corresponding author: Yong-Jin Kim (kyj0916@126.com).

**ABSTRACT** In recent years, cloud computing has been developing rapidly and is widely used in various fields such as commerce and scientific research. However, security issues, including access control, are a very important problem in popularizing cloud computing and this has influenced its wide application of cloud computing. As one of the solutions to these problems, we have proposed a blockchain-based role-based access control model with the separation of duties constraints in a cloud environment. In the model, we used Hyperledger Fabric as a blockchain platform for storing the access control policies and provided several functions for effective role management. In addition, we presented an access control scheme for cloud storage data by combining the proposed model and the verification mechanism for the user's ownership of a role and analyzed the security properties of the scheme. Finally, we deployed Hyperledger Fabric test network, implemented an online test system that performs access control using the proposed scheme in the Ali cloud environment, and evaluated the model performance in this scenario.

## I. INTRODUCTION

Cloud computing refers to both the applications delivered as services over the Internet and the hardware and system software in the datacenters that provide those services [1]. For many enterprises, it is difficult to determine the enterprise scale in the early stage of business to meet the changing needs of users, which will affect the scale of hardware resources required for enterprise operation. Even if the company has constructed its hardware center, it will spend a lot of money on the purchase and maintenance of these resources. Cloud computing can help these companies to reduce the basic investment in hardware centers by providing hardware resources with limited resources according to users' demands that change in real time; thus, many companies prefer cloud services by cloud service providers such as AWS, Google, and Microsoft. However, security issues, including access control, have become a very important problem in popularizing cloud computing, influencing its wide application of cloud computing.

Blockchain technology was first introduced in 2008 by Satoshi Nakamoto with the advent of Bitcoin, which is a peer-to-peer electronic cash system. A blockchain can be defined as an immutable ledger for recording transactions maintained within a distributed network of mutually untrusting peers [2]. As the Ethereum blockchain supports a smart contract, a Turing-complete language, which appeared in 2013, blockchain is being used not only for trust exchange, but also as a basic technology to ensure safety in healthcare management, IoT, and other business processing fields. Blockchains are classified into permissionless and permissioned blockchains. In a permissionless blockchain, the content of the distributed ledger is shared with the public and anyone can participate in the decision-making process. In a permissioned blockchain, data are shared only between specific organizations or consortiums. Both Bitcoin and Ethereum are included in a permissionless blockchain.

Hyperledger Fabric used in this paper is a permissioned blockchain. Hyperledger Fabric, or Fabric for short, is the contribution made initially by IBM and Digital Assets to the Hyperledger project [3]. Fabric enables blockchain networks to be built using a modular, open, and flexible approach. Compared to the permissionless blockchain, Fabric does not require a mining process; therefore, it provides high system performance and supports scalability, privacy, and confidentiality, making it a type of blockchain that is widely used in business processes.

Access control is a technology that restricts subjects from accessing resources according to rules prescribed within the organization, and various access control models have been developed thus far, including discretionary access control (DAC), mandatory access control (MAC), role-based access control (RBAC), and attributed-based access control (ABAC).

The DAC assigns access permissions to subjects based on their identities. It is used in many operating systems including Windows and Unix. However, in organizations with many subjects and resources, access control enforcement is not effective, and a lot of storage space is required to store assignment relationships between subjects and permissions; therefore, it is not suitable for these organizations.

In MAC, access control is enforced by the system security administrator, and the resource owner is not allowed to modify the access control policy. Owing to the simple structure of permissions and strict enforcement of access control, it is widely used in the military field.

In ABAC, access control is enforced based on attributes of the subject and context. ABAC is a powerful, flexible, and widely used access control model, but in a trans-organizational environment and in an environment where changes to the job, such as job addition and deletion, frequently occur within the organization, implementation of ABAC is complex, and ABAC is not suitable in such an environment. This problem can be solved using RBAC. RBAC can effectively and flexibly manage the assignment between the subject and permission by introducing the concept of a role. The flexibility and ease of management are the biggest advantages of RBAC, which is widely used in industrial and commercial application systems.

The separation of duties (SoD) is a powerful constraint for implementing the concept of least privilege and avoiding one-man control [4, 5]. The purpose of separating duties in RBAC is "to ensure that failures of omission or commission within an organization are caused only as a result of collusion among individuals. To minimize the likelihood of collusion, individuals with different skills or divergent interests are assigned to separate tasks required for the performance of a business function. The motivation is to ensure that fraud and major errors cannot occur without deliberate collision of multiple users" [6]. To avoid one-man control, the business function or process is divided into multiple roles, which are then assigned to different users so that no single user can activate all the roles to avoid the execution of a business function or process [7]. Therefore, considering the SoD constraint in access control enforcement is essential for a higher system safety assurance.

In recent years, many studies introducing the implementation of the blockchain-based RBAC model have been published; however, few attempts have been made to implement the SoD constraint. Some articles mentioning the SoD constraint considered only the SSoD and not the DSoD. In addition, in these articles, access control performance was limited because public blockchain was used, and a method to provide both role-based access control and verification for role ownership has not yet been proposed. Therefore, in this paper, we proposed a blockchain-based RBAC model with SoD constraints and provided an access control scheme with higher security in a cloud environment using the model.

The main contributions of this paper are as follows:

1. In this paper, we proposed a blockchain-based access control model that can effectively prevent policy tampering attacks and provide strict enforcement of access control by considering SoD constraints while storing access control policies using blockchain technology.

2. We presented an access control scheme for cloud storage data by combining the proposed model and the verification mechanism for the user's ownership of the role, the prevention of the role forgery attack can be ensured at a high level.

3. We deployed Hyperledger Fabric test network, implemented an online test system that performs access control using the proposed model in the Ali cloud environment, and performed a model evaluation in this scenario.

The rest of this paper is organized as follows. We first introduced recent research on blockchain-based RBAC models and RBAC models with SoD and compared these models for several properties in Section 2. In Section 3, the architecture of the RBAC model and the implementation of the model on the blockchain are proposed along with the access control scheme in a cloud environment. In Section 4, the security analysis of the proposed model and performance evaluation results in an online test scenario are presented. Finally, we discuss our conclusions and future work.

## II. RELATED WORKS

In recent years, many studies have been published on the implementation of blockchain-based RBAC models and RBAC models with SoD. David W Chadwick *et al.* [8] provided a mechanism for implementing DSoD constraints in a multi-session environment by constructing a hierarchical structure based on business context and defining mutually exclusive roles (MER) set according to business context and Muhammad Asif Habib *et al.* [7] overcame the problems raised in the implementation of DSoD constraint at the role level by implementing the model at the permission level while considering the conflicting permission set. The authors provided three modes for the implementation of DSoD constraints so that the model could be used in different situations according to the requirements for safety.

Yi Ding *et al.* [9] presented an access control model called SC-RBAC using smart contracts in a DApp environment. The authors defined an administrative role in the model and used it to manage the access control policy of the application. They also defined three smart contracts for managing resources, roles, and users of the model and implemented the simplest core RBAC. This paper does not effectively design the storage and management of access control policies and does not consider constraints at the enforcement time, which can raise the security and efficiency problems of access control. Therefore, it is not suitable for organizations with many roles and users.

JASON PAUL CRUZ *et al.* [10] discussed a role authorization method between a role-providing entity and a service-providing entity based on the Ethereum blockchain for implementing the RBAC mechanism in a trans-organizational environment. The authors presented a verification workflow of the user's ownership of a role based on the user's private and public key pair using a smart contract and challenge-response protocol. The main contribution of this paper is to propose a verification method for the user's ownership of the role but the detailed implementation of access control enforcement is not discussed.

Mohsin Ur Rahman *et al.* [12] proposed the RBAC implementation model based on blockchain in a decentralized online social networks environment. Along with core RBAC, the most basic RBAC model, verification of the user's ownership of the role is also implemented using the key pair of the user's private key and public key. Mohsin Ur Rahman *et al.* [13] increased the throughput of the system by implementing the RBAC model on the EOS blockchain rather than on Ethereum. The authors provided model safety and convenient management of access control policy to some extent by considering the features of role hierarchies, constraints of the subject-role assignment, and role-permission assignment while implementing the RBAC model on the blockchain. The constraint principle in the user-role assignment is defined using a predicate. Considering constraints can be seen as an advantage of this paper but the authors considered the constraint only in design time and not in runtime.

Danyang Liu *et al.* [14] verified the data user's ownership of the role through identity-based signature using Ethereum account public and private keys. In addition, the data owner uses a hierarchical attribute-based encryption algorithm to encrypt the access structure information of the model and validates the data user's access request by calling the method defined on the blockchain. This paper focuses on the verification of the user's ownership of the role and the detailed implementation of policy enforcement is not discussed. Table 1 below shows the comparison results of these papers for several attributes.

TABLE I
THE COMPARISON RESULTS OF THE PAPERS INTRODUCING THE RBAC MODEL FOR SEVERAL PROPERTIES

| Papers | Blockchain Platform | Role ownership verification | Role hierarchy | SSoD Constraint | DSoD Constraint |
|---|---|---|---|---|---|
| [8] (2007) | X | No | No | No | Yes |
| [7] (2014) | X | No | No | No | Yes |
| [9] (2019) | Ethereum | No | No | No | No |
| [10] (2018) | Ethereum | Yes | No | No | No |
| [11] (2019) | Ethereum | Yes | No | No | No |
| [12] (2019) | Ethereum | Yes | No | No | No |
| [13] (2020) | EOS | No | Yes | Yes | No |
| [14] (2021) | Ethereum | Yes | Yes | No | No |
| Our paper | Hyperledger Fabric | Yes | Yes | Yes | Yes |

Yong Joo Lee *et al.* [11] used "Role Issuer" and "Role Verifier" which are responsible for role issuance and role verification to provide role-based access control with user authentication. The access requestor sends the access request through the hash value of the combined data of its Ethereum public address and the role ID issued by the role issuer. In this way, the authors implemented anonymity-based user authentication.

## III. IMPLEMENTATION OF THE PROPOSED MODEL IN CLOUD ENVIRONMENT

In this section, we introduce the application of the proposed model to the cloud environment. First, the architecture of the proposed model and the detailed implementation of the model on the blockchain are introduced, and then the access control scheme steps for handling users' access requests are discussed.

### A. DESIGN OF THE PROPOSED MODEL

The proposed model for access control consists of several components including "User," "Role," etc. Fig.1 shows the interrelationship between these components and the architecture of the proposed model.

#### 1) KEY COMPONENTS OF THE MODEL

In the proposed model architecture, "Object" means the organization's resources, and each object is identified using Object ID. "User" is the subject requesting access to the object. The model also includes the other components including "Role" and "Constraint for SSoD and DSoD principle" and these components are introduced below in detail.

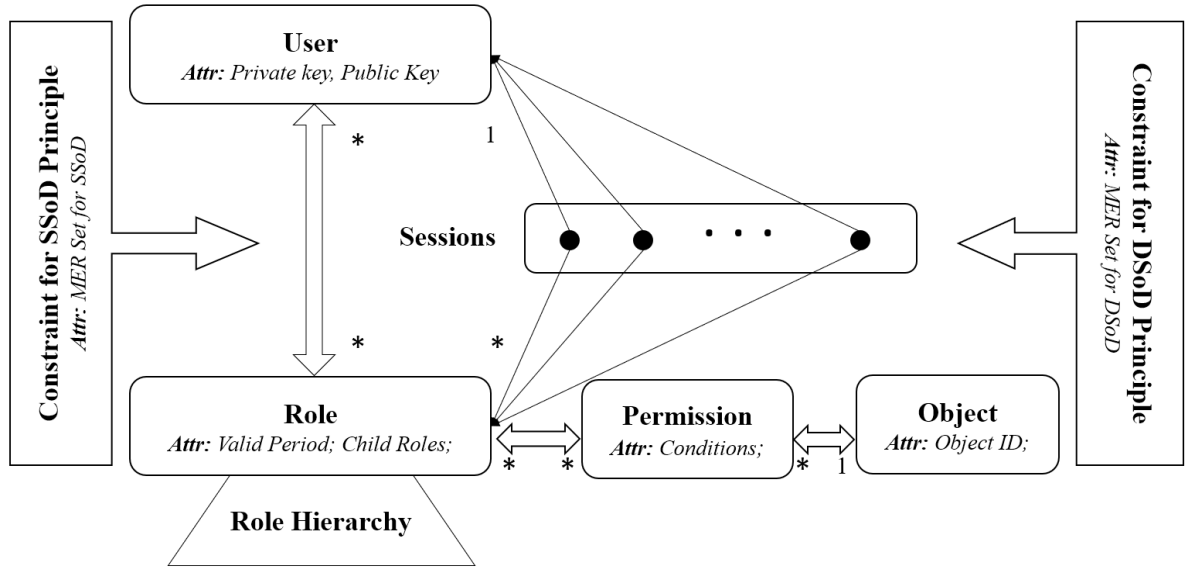

**FIGURE 1.** The architecture of the proposed model

*a: ROLE CONFIGURATION*

The user first generates a key pair of private and public keys on the local computer. The private key is used to verify the user's ownership of the role when requesting access to the object and the public key is used for user identification. After generating the key pair, the user sends their public key and required role to the data administrator. After receiving the user's request, the data administrator first checks whether assigning the required role to the user conforms to the SSoD principle specified by the organization based on the user-role assignment information stored on the blockchain. If the check is successful, the user-role assignment information is updated and the result is returned to the user. The user-role assignment information is stored using a map data structure with the public key of the user as the key and the array of roles assigned to the user as the value. Each specific role consists of the following data fields:

*Valid period*: This field stores the value of the maximum period that the role can be assigned to a user. The valid period is checked after verifying the user's ownership of the role. This check is performed using the history of the user's request event stored on the blockchain. When handling an access request, if the valid period has already elapsed from the time the role is assigned to the user, the request is denied and the request event is recorded on the blockchain.

*Child roles*: This field stores the role identifiers of all child roles of this role which represent the role hierarchy relationship. The concept of the child role is introduced in detail in section B below.

*Assigned permissions*: This field stores permissions assigned to the role. Each permission entity consists of Object ID, available operations on the object, and conditions that must be satisfied to operate on the object. For example, the permission entity ("Prob1-1," "W," "st:2021-12-22 15:00:00, ed:2021-12-22 15:40:00") indicates the write operation to the "Prob1-1" object and this operation is available from "2021-12-22 15:00:00" to "2021-12-22 15:40:00". The condition can be information such as the user's location, public key, access history. If there is no condition, the condition parameter is set to "null".

*b: CONSTRAINT*

The concept of constraint emerged after the RBAC model was proposed. Constraint indicates the consideration of various conditions such as restriction to user-role assignment and role-permission assignments when enforcing access control policy and can help provide strict access control enforcement. The constraints in the RABC model are classified into context and authorization constraints, which are discussed in detail below.

*i) Context constraint*

The content of context constraint is to enforce access control based on the context (request time, location, user's nationality, etc.) which is independent of the model's internal state when users request permissions. There are two ways to deal with dynamically changing contexts. One is to modify the role-permission assignment according to the context, and the other is to specify the conditions to operate for each permission. The access control policy is frequently modified in the real world and the performance of storage operation is not high enough on the blockchain, so the first one is not suitable for the blockchain-based access control model, and conditional permission is introduced in our model. The content of conditional permission is introduced in detail above.

*ii) Authorization constraint*

Unlike the context constraint, the authorization constraint is enforced based on the internal state of the access control model. SoD is the most representative example of authorization constraint. It is classified into the static separation of duties (SSoD) constraint and the dynamic separation of duties (DSoD) constraint [15]. We specified the SoD principle through the definition of MER set in the proposed model and a user cannot activate multiple roles specified in the MER set simultaneously. Taking the online test system as an example, one user cannot own both the "Reviewer" role and "Student" role at the same time according to the system requirements for safety. In addition, a user can own but cannot activate the both "Reviewer" role and the "Editor" role at the same time in one session. We can express the MER set for this SoD principle as *MER* ({*Reviewer*, *Student*}, *2*, *"Static"*) and *MER* ({*Reviewer*, *Editor*}, *2*, *"Dynamic"*).

The SoD principle can be stored on the blockchain using both the array data structure of MER roles and XML language. To describe the organizational policy for access control, much research has been conducted on various languages, such as RCL and XML. The expression language of the principle should be free from ambiguity with powerful expressions and should be easily understood by humans and computers. XML is highly expressive and consumes less memory for storing policy strings; therefore, we used the XML language to express the SoD principle.

Each method of using an array data structure and XML language has its drawbacks. If the MER set is stored using an array data structure, the user's access request can be verified in a short time and consumes less storage space on the blockchain; however, it is difficult for people to understand the policy. Using XML language, people can easily understand the policy, but to verify the user's access request, we need to interpret the XML language expressing the SoD principle which can lead to a decrease in the model performance and consume a lot of storage space for principle storage. Fig.2 shows an example of expressing SoD principle using XML language.

Therefore, the model provides a method for storing access control rules using an array data structure on the blockchain in addition to XML. Therefore, we provided both methods so that they can be used according to different needs. The data administrator calls the corresponding method on the blockchain to store the principle expressed in an array data structure or XML language. Once the policies are stored on the blockchain, the CSP verifies the user's access request by referring to the data.

### 2) FUNCTIONS FOR ROLE MANAGEMENT OF MODEL

In addition to the components introduced above, the model provides other functions including delegation, revocation, and normalization of role hierarchy for efficiency and safety of the access control. These functions are introduced below in detail.

*Delegation* is a process in which a user transfers some permissions of a role assigned to him/her to another user; it is a useful feature in many cases. Taking the online test system as an example, a user who has the reviewer role may not be able to review students' test answers due to illness or other special reasons, which creates an obstacle to the normal operation of the system. In this case, the user can transfer the review authority he/she owns to another user he/she trusts for a certain period to replace his/her duty, and the system operation will not be affected. In this way, delegation enables the normal operation of the system to be maintained even when the conditions for the user to exercise authority are not satisfied. This feature is useful not only for online test systems but also for many commercial organizations. We express delegation manipulation using information such as delegator's ID, delegate's ID, delegated permissions, and delegation period.

```
<SoDPrinciple org="OnlineTest">
    <MERSet type="Static" cardinality="2">
        <Role value="Reviewer">
        <Role value="Student">
    </MERSet>
    <MERSet type=" Dynamic" cardinality="2">
        <Role value="Reviewer">
        <Role value="Editor">
    </MERSet>
</SoDPrinciple>
```

**FIGURE 2. Example of expressing a SoD principle using XML language**

*Revocation* is the process of revoking a previously assigned role from a user. In the real world, users are not allowed to permanently own their assigned roles according to the safety requirements of the organization. From this, the revocation has become the essential feature of the access control mechanism, so we specified a valid period of role-owning for each role in the proposed model, and role revocation is automatically performed when the role's expiration date is reached.

The *normalization of role hierarchy* is the process of generating a role hierarchy from the role-permission assignment. Hierarchical RBAC provides for the establishment of role hierarchies, with senior roles adopting all the permissions within junior roles [1]. The data administrator specifies the organization's access control policy in a manner that defines roles and their corresponding permissions. In this case, an inclusion relationship can exist between permission sets assigned to roles. For example, in an online test system, there is a top reviewer, a job that conducts a final check on all review results. The top reviewer has all the permissions that reviewer-1, a job that reviews only Question 1, and reviewer-2, a job that reviews only Question 2, has. In this case, the top reviewer role becomes the parent role of reviewer-1 and reviewer-2. In the normalization algorithm, we first defined the inclusion relationship between roles based on the permission sets assigned to the roles. Based on this relationship, we created a graph with each role as a vertex. Then, a normalized role hierarchy is generated from this graph using the DAG normalization algorithm [16]. Through the normalization algorithm, the hierarchical structure between roles is generated from the role-permission assignments specified by the data administrator. The application of role hierarchy in the RBAC model can help not only simplify of the administration of the model but also reduce the space for storing the access control policy. Unlike general databases, blockchain provides only limited data storage space, so this function to reduce the storage space consumption has a positive effect on the system performance improvement.

We installed chaincodes for delegation, revocation, and normalization of the role hierarchy on Hyperledger Fabric and updated the information of user-role assignment and role configuration stored on the blockchain in these chaincode methods.

### B. MODEL IMPLEMENTATION ON HYPERLEDGER FABRIC

In this section, we introduce the methods for implementing the model introduced above on Hyperledger Fabric. To implement the model, several chaincode methods for the configuration of the model, role management of the model, and handling requests using the model are implemented.

#### 1) CONFIGURATION OF THE MODEL

In this section, we introduced the methods used to configure our model. These methods are called to manage access control policies that are composed of the role-permission assignment and SoD constraint. These methods can also be called to record the user's access request history which can be used as information for providing role revocation, attacker tracking, and implementation of the SoD constraint. Methods used to configure our model are as follows:

*SetRoleConfiguration* (*Role ID, Assigned Permissions, Valid Period*): Set the valid period of a role along with the assigned permissions for each role. Each permission is a conditional permission (described in detail in Section A of 3.1) and consists of information such as object ID, operations to the object, and operating conditions. The valid period is the maximum period that a user can own a role, and role revocation is performed automatically once a specified time has elapsed after role authorization. The role-permission assignment information is stored using a map data structure on the blockchain. Only a data administrator can call this method.

*SetSoDConstraint* (*MER set*, *k*, *Type of the constraint*): Store the "MER set" on the blockchain to implement the SoD constraint in the organization. Using the parameter "k," the maximum number of the roles that can be authorized or activated simultaneously are specified, and "Type of the constraint" specifies whether the "MER set" is for SSoD constraint or DSoD constraint. Only the data administrator can call this method.

*AppendReqRoleHistoryEntity* (*User's Public Key*, *Required role*): Store the record of the user's role authorization request event on the blockchain. Each history entity includes information of the requestor's public key, request time, required role, and request result. For example, we can express the user's access request history entity as (0X3565CE13E9F0AEBA0AB5A03615EA2134, "2021-12-21 17:37:00," "Student," "Allowed"). Only the CSP can call this method.

*AppendReqHistoryEntity* (*User's Public Key*, *Object ID*, *Required Permission*, *Request Result*): Store the record of the user's access request event on the blockchain. Each history entity includes information of the requestor's public key, request time, Object ID, required permission, and request result. For example, we can express the user's access request history entity as (0X3565CE13E9F0AEBA0AB5A03615EA2134, "2021-12-21 17:37:00," "Prob1-1," "Write," "Denied"). Only the CSP can call this method.

#### 2) ROLE MANAGEMENT OF THE MODEL

In this section, we introduce the methods for implementing role management of the model introduced above. The role management of the model includes the functions of role revocation, delegation, and the normalization of the role hierarchy. These methods are called by a data administrator to perform role management. The CSP can perform role revocation when handling a user's access request.

*RoleRevocation* (*User's public key*, *User's Role*): This method is called by the CSP if the valid period for role ownership is exceeded when validating a user's access request to the resource or by the data administrator when a change occurs in access control policy within the organization. The user-role assignment information stored on the blockchain is updated when this method is called. This method can also be called by a data administrator when a user's strange behavior is detected.

*SetDelegation* (*Delegator's Public Key*, *Delegate's Public Key*, *Delegated Role*, *Expiration Time*): This method is called to perform the delegation operation and the public keys of delegator and delegate, the delegated role, and the expiration time of the delegation are input as parameters. The information of this delegation operation is stored using the array data structure on the blockchain is used to verify whether the user's request to activate the role can be accepted.

*NormalizeRoleHierarchy* (): Generates the role hierarchy from the role-permission assignment information specified by the data administrator. In this method, we normalized the role hierarchy by utilizing the DAG normalization algorithm on the graph, which is generated based on the inclusion relationship between permission sets assigned to each role. Subsequently, role IDs of child nodes are appended to the "Child roles" field of each role based on the normalized role hierarchy.

#### 3) REQUEST HANDLING USING THE MODEL

After receiving a role authorization request or resource access request from users, these methods validate the request using the proposed model and then return the result to the user.

*RequestRoleForUser* (*User's Public Key*, *Required Role*): This method performs role issuing after receiving a role authorization request. Only a data administrator can call this method. In this method, we first record the request event on the blockchain and check whether the required role can be assigned to the user based on the SoD constraint information. If the check is successful, the user-role assignment information stored on the blockchain is updated and the check result is returned to the user.

*RequestAccessToRes* (*User's Public Key*, *Required Role*, *Required Operation*): This method checks whether the user's request can be accepted after receiving the access request to the resource. Only the CSP can call this method. In this method, we first record this request event on the blockchain and check whether the required role is assigned or delegated to the user using the information of the user-role assignment and delegation stored on the blockchain. If the checking is successful, we verify the user's ownership of the role, the expiration date of the required role, and the DSoD constraint. Subsequently, we return the checking result to the user. The workflow of this method is described in detail in the following section.

### C. ACCESS CONTROL SCHEME IN CLOUD ENVIRONMENT

In this part, we introduced the access control scheme for handling the users' access requests using the proposed model. The user's access request is handled through interaction between the CSP, data administrator, and Hyperledger Fabric.

#### 1) INTRODUCTION TO THE SCHEME COMPONENTS

The scheme components consist of the CSP, data administrator, and Hyperledger Fabric. The functions of these three components participating in user's access request handling are as follows.

*a: CLOUD STORAGE PROVIDER*

The main function of the CSP is to handle the user's access request to the resource.

After receiving the user's request, the CSP records the request event on the blockchain. The first process of handling is to check the expiration time of the role assigned to the user. This check is performed using the information of the user's role-issuing time stored in the user's event history on the blockchain, and the information of the valid period in the role.

Subsequently, the CSP verifies the user's ownership of the role. It randomly generates a message which will be used to verify the user's ownership of the role and then sends it to the user. Once the user receives the message, they generate the message signature using their secret key and send it back to the CSP. The CSP verifies the message signed with the user's public key, and if the verification is successful, the DSoD constraint is checked using the information of the user's session and the MER set specified by the data administrator.

*b: DATA ADMINISTRATOR*

The data administrator performs the following two functions.

First, it is responsible for managing the access control policy. It specifies the access control rules within the organization and defines the roles, permissions, and role-permission assignment relationships based on information about the jobs and the permissions required to perform the job. Simultaneously, it creates and deletes the databases used for the organization's operations.

Second, it performs role registration for the user based on the organization's SSoD constraints. Simultaneously, it records the user's request event and stores the user-role assignment information which will be used to handle the user's access request on the blockchain.

*c: HYPERLEDGER FABRIC*

Hyperledger Fabric is responsible for the following three functions.

First, Fabric is used to store the organization's access control policy specified by the data administrator. The information of the user-role assignment and the role-permission assignment which will be used to enforce access control is stored on it. The information is updated by the data administrator when it handles the user's role registration request and the organization's access control policy is changed.

Second, Fabric is used to store the user's request event history for audit and constraint implementation. Once a suspicious behavior is detected in the model, a malicious user can be tracked using the history stored on the blockchain; at this time, it is almost impossible to modify the history information stored on the blockchain. Each history entity includes the user's public key, request time, object ID, required permission, and the response result. The request history can be used not only for audits but also for constraint implementation. After receiving the user's request for role activation, the CSP determines whether the request can be allowed using the information of the DSoD constraint and the user's request history stored on the blockchain. We can also use history information to handle a user's access request to the resource. This type of access control is called history-based RBAC. Because this content is beyond the scope of this paper, it is not specifically introduced here.

#### 2) SCHEME EXPLANATION

Before receiving the user's access request, the data administrator first specifies the organization's access control policy, stores it on the blockchain, and then handles the user's role registration request based on the SSoD principle. If role registration is successful, the user can send an access request to the resource. After receiving the user's request, the CSP checks whether the user is the correct owner of the role by verifying the message signature created by the user using the user's private key. The CSP then verifies the valid period of the role and checks whether the DSoD principle is satisfied based on the information of the user's request event history stored on the blockchain. If the check is successful, the user's request event is recorded on the blockchain and the result is returned to the user. The detailed steps of the scheme are illustrated in Fig. 3.

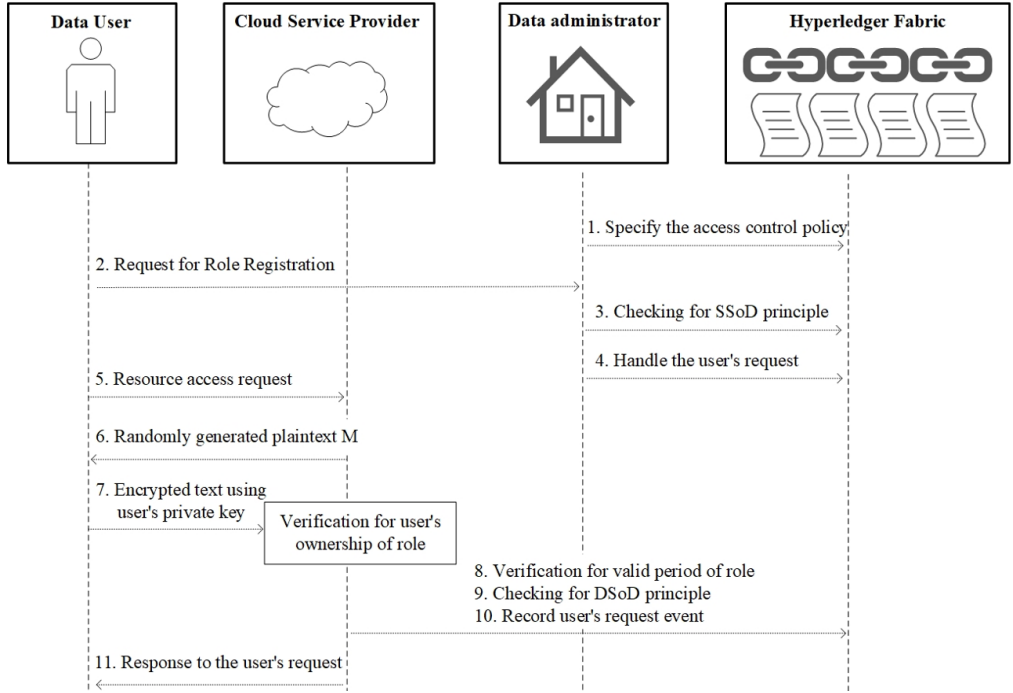

**FIGURE 3.** The access control scheme in a cloud environment

## IV. MODEL EVALUATION

Security and performance are key factors in the evaluation of access control models. In this section, we analyze the security properties provided by the proposed model architecture and evaluate its performance in an online-test system scenario.

### A. SECURITY ANALYSIS

The implementation of RBAC considering the SoD on the blockchain provides security for access control in various aspects.

#### 1) PREVENTION TO THE POLICY TAMPERING ATTACK

A policy tampering attack is one of the most common and dangerous attacks on access control. In previous server-centric models, the organization's access control policy is stored in the central server, so the malicious users can easily achieve the purpose of modification of the policy. In addition, the records of users' request events are stored in the server; therefore, once the malicious user deletes or changes the records, it is almost impossible to track the attacker. In this model, we can overcome these obstacles using blockchain technology. Blockchain only supports the appending operation of the data, and deletion or modification of the data is not allowed. Owing to the characteristics of blockchain technology, only attackers with sufficient computational ability can delete or modify recorded data. The sufficient computational ability mentioned here indicates the ability to exceed the total computational power of legitimate users constituting the system, and it hardly exists in the real world. Thus, once the policy is stored on the blockchain, it is almost impossible to modify it. By storing not only the policy but also the history of the user's request events on the blockchain, safety for history can be ensured, and this information can be a powerful means to support the tracking of the attacker.

Overall, the management of the access control policy using Hyperledger Fabric can prevent the policy modification attack at a high level.

#### 2) PREVENTION TO THE ROLE FORGERY ATTACK

In the RBAC model, access control is enforced based on the role information assigned to users; thus, providing accurate information about the user's ownership of the role is an important requirement to ensure the security of the access control. In a real environment, attackers can provide forged information about the user-role assignment to the service provider in various ways, which threaten the normal operation of the system. Thus, the verification of the user's ownership of the role is required; therefore, we introduced a verification mechanism along with the design of the proposed model. The user proves that he/she is the correct owner of the role through interaction with the service provider, and the key pair of the private key and public key he/she generated is used in this process. Because the user's private key is created by the user and is known only to the user, an attacker cannot forge information about the user-role assignment.

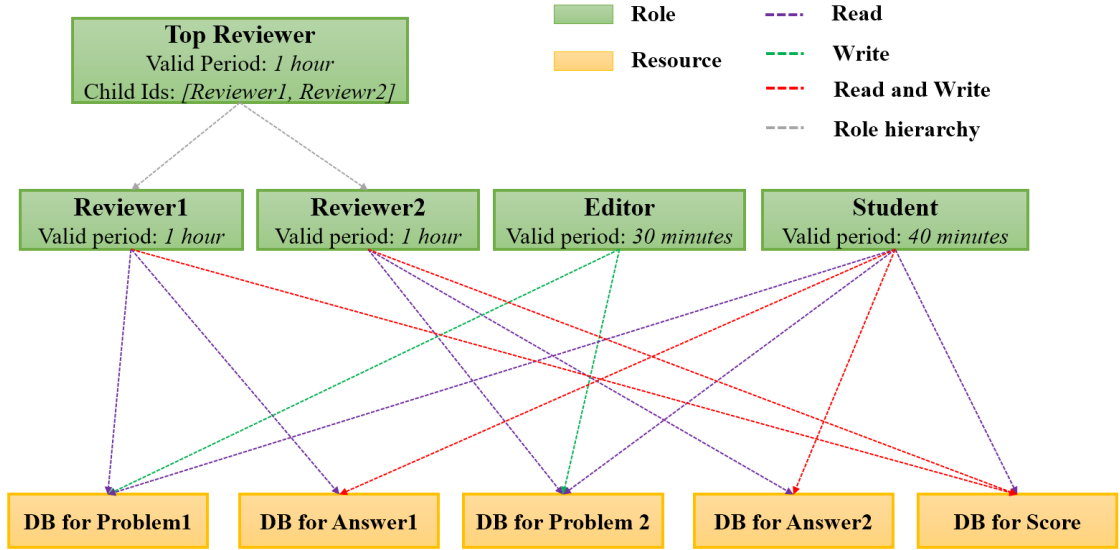

**FIGURE 4.** Illustration of the role hierarchy and role-permission assignment

At the same time, the verification is performed using plain text randomly generated by the service provider each time, so even if the attacker captures the packet exchanged during the verification process, the attacker cannot reuse the captured packet for the next attack. Thus, replay attacks can be effectively prevented.

*3) STRICT ENFORCEMENT OF THE ACCESS CONTROL POLICY*

The principle of least privilege and separation of duties are essential features for strict enforcement of the access control policy in organizations. In this paper, a role-based access control model is used and permissions required for users to perform their jobs are assigned to roles that a user owns. By using this role, unlike other general access control models, users can only own the permissions necessary for performing the job and cannot own more permissions, and the least privileged feature of access control can be implemented effectively at a low cost.

In addition, we designed a mechanism for the implementation of the SoD constraint and used the MER set specified by the data administrator. Considering the SoD constraint at design time and runtime prevents users from owning conflicting permissions and ensures that permissions are not misused. Overall, we provided strict enforcement of the access control policy at a high level by considering these two features.

*B. PERFORMANCE EVALUATION*

In this section, we evaluate the performance of the proposed model. We implemented an online-test system using the proposed access control scheme in the Ali cloud environment and measured the average response time for each user's access request in this scenario.

*1) USE-CASE SCENARIO FOR THE EVALUATION*

A testing scenario for an online test system is designed to evaluate the proposed model. An online test that does not use paper is called a computer-based test (CBT) or computer-assisted test (CAT). All test processes are done using computers, ranging from constructing questions, classroom settings, user settings that can take the exam, setting teachers teaching, until the examination process, by utilizing web engineering technology [17]. The online test provides convenience by allowing students and reviewers to browse and review test questions at the test center configured on the Internet without moving. Next, unlike the general test, the online test can prevent various negative phenomena because the test answers and scoring results are transmitted directly from the student to the reviewers and from the reviewers to the students without going through an intermediate stage. In addition, it has the advantage of reducing the cost of resources including paper used for the test. Because of these advantages, online test systems are widely used in various fields including education. We evaluated the performance of the model by simulating an online test system using the proposed model in a cloud environment.

We defined five roles and five resources for this scenario. Fig. 4 illustrates the role hierarchy and role-resource assignment.

The meaning of each element in Fig. 4 is as follows.

*DB for Problem1/DB for Problem2* (Object): A database of Problem 1 and 2 presented in the test.

*DB for Answer1/DB for Answer2* (Object): A database of the students' answers to Problem 1 and 2.

*DB for Score* (Object): A database of the reviewer's scoring results.

*Reviewer1* (Role): As a role with scoring authority for Problem 1, it has "read" permission to the database of Problem 1 and students' answer to Problem 1, and "write" permission to the score database. In this scenario, the review time for the answer is limited to 1 h; therefore, the valid period for this role is set to 1 h.

*Reviewer2* (Role): Has the same authority as "Reviewer1". The difference is that they have scoring authority for Problem 2, not Problem 1.

*Top reviewer* (Role): As a role that re-examines the scoring results, it has the authority of both the "Reviewer1" role and the "Reviewer2" role. The valid period for this role is also 1 h.

*Editor* (Role): As a role with editing authority for the problem, it has "read"/"write" permission to all databases of problems. The valid period for this role is 30 minutes.

*Student* (Role): As a role corresponding to the student taking the test, it has "read" permission for all databases of problems, and scoring results, and "write" permission for the answer database. The valid period for this role is equal to the total testing period. In this example, this is set to 40 min.

### 2) TEST ENVIRONMENT CONFIGURATION

For the evaluation, we installed an Alibaba Cloud Linux 3.2104 containing 40GB SSD, 2GB RAM, and 1MB bandwidth and Hyperledger Fabric 2.0 on the Ali cloud server. We created a Fabric test network that consists of two peers, one ordering node, using scripts provided in the fabric-sample repository. Peers store the blockchain ledger and validate transactions before they are committed to the ledger. An ordering service allows peers to focus on validating transactions and committing them to the ledger. For simplicity, a single-node Raft ordering service is configured. To reduce complexity, a TLS certificate authority (CA) is not deployed. All certificates are issued by the root CAs. The sample network deploys a Fabric test network using the Docker Compose tool. Because the nodes are isolated within a Docker Compose network, the test network is not configured to connect to other running Fabric nodes. Hyperledger Fabric Docker images are downloaded from https://goproxy.io for the configuration of the test network. We use the Go language to write the chaincode and create an evaluation project for the proposed model. Fig.5 below shows the creation result of the Fabric test network.

```
Creating network "fixtures_test" with the default driver
Creating volume "fixtures_orderer.example.com" with default driver
Creating volume "fixtures_peer0.org1.example.com" with default driver
Creating volume "fixtures_peer1.org1.example.com" with default driver
Creating couchdb1             ... done
Creating orderer.example.com  ... done
Creating couchdb0             ... done
Creating ca.org1.example.com  ... done
Creating peer1.org1.example.com ... done
Creating peer0.org1.example.com ... done
```

**FIGURE 5.** Result of deploying Fabric test network

### 3) EVALUATION RESULT

We evaluated the performance of two features of the model. The first is the total response time of the system for requests from different amount of users, and the other is the total response time for the enforcement of the SoD constraint. To simulate the simultaneous requests sent by multiple users in a cloud environment, sending an access request to resources and handling a user's request are performed in goroutines that are created and executed in parallel. Because the state of the machine installed in the cloud varies over time, it is not possible to evaluate the exact performance with one measurement; therefore, we performed the measurement 10 times and recorded the average value as the result. Before measuring the response time, the role registration is performed for each user in the scenario. When an access request is received, the user's ownership of the role is verified, and the chaincode method for request handling installed on the blockchain is called in a goroutine. We defined the time interval between the creation time of the goroutine that handles the request for the first user and the finish time of the goroutine that handles the request for the last user as the response time and measured this value.

To evaluate the first feature, we assume that the number of students in one class to 30 in the online test and measured while changing the number of system users from one class to seven classes. Fig. 6 shows the relationship between the number of system users sending access requests and the total response time to the requests. From Fig. 6, it can be observed that the total response time has a linear relationship with the number of model users. The average response time of the system for each user is approximately 55ms, and the number of user requests processed per unit second is 20. Therefore, the proposed model can provide sufficient performance necessary for the operation of an organization in a real environment.

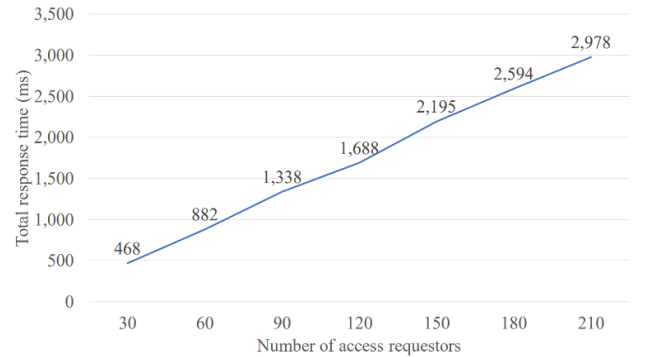

**FIGURE 6.** Total response time according to changes in the number of access requestors

Fig. 7 shows the response time of the proposed model for the enforcement of the DSoD constraint when the number of malicious users requesting conflicting roles changes from 10 to 100 for 100 system users.

For the measurement, we first specified the MER set for the DSoD constraint of the test scenario by calling the chaincode and then activated a role for each user. Next, for the number of users to be measured, an activation request to the role that conflicts with an already activated role is sent, and the total response time is recorded. As shown in Fig. 7, the total response time is almost constant in each case, and from this, we can conclude that the system can keep almost stable performance for different numbers of malicious users. The enforcement of the SSoD constraint yielded similar performance evaluation results.

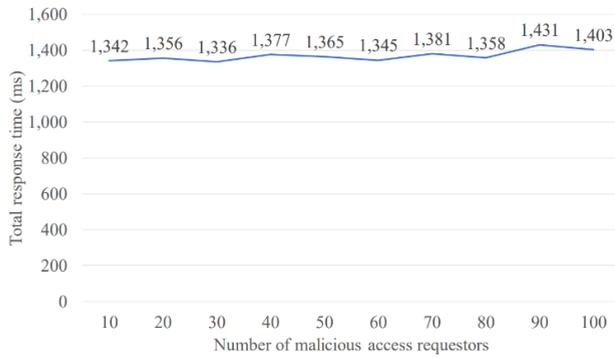

**FIGURE 7. Total response time for the enforcement of SoD constraints**

## V. CONCLUSION

In this paper, we presented a blockchain-based RBAC model with SoD constraints and an access control scheme in a cloud environment using the model. In our model, both the SSoD and DSoD constraints are implemented by specifying the organization's MER set on the blockchain, and functions for role management are implemented on the blockchain. In the proposed access control scheme, the user's ownership of a role is verified using a keypair of private and public keys generated by the user, and the access request is handled using our model. Finally, we presented the security analysis of the proposed scheme and evaluated its performance. After simulating an online test system scenario using the proposed scheme in the Ali cloud environment, the performance of the model according to the change in the number of system users and malicious access requestors is evaluated, resulting in an average response time for each user is about 55ms.

In future work, we will improve the model architecture to achieve lower response time and scalability. In addition, in the proposed model, the SoD constraint is implemented only for a single-session environment, but the SoD constraint for a multi-session environment will also be considered in the future.

## VI. ACKNOWLEDGMENT

We thank the endorsements, associate editor, and Editor-in-Chief for their valuable feedback on the paper.